# A Response-Function-Based Coordination Method for Transmission-Distribution-Coupled AC OPF

Zhengshuo Li
Tong Xu
Tsinghua-Berkeley Shenzhen Institute (TBSI)
Shenzhen, Guangdong, 518055, China
shuozhengli@sina.com

Qinglai Guo
Hongbin Sun
Dept. of Electrical Engineering
Tsinghua University
Beijing, 100084, China
shb@tsinghua.edu.cn

Jianhui Wang
Department of Electrical Engineering
Lyle School of Engineering
Southern Methodist University
Dallas, TX 75275, UA
jianhui@smu.edu

*Abstract*—With distributed generation highly integrated into the grid, the transmission-distribution-coupled AC OPF (TDOPF) becomes increasingly important. This paper proposes a response-function-based coordination method to solve the TDOPF. Different from typical decomposition methods, this method employs approximate response functions of the power injections with respect to the bus voltage magnitude in the transmission-distribution (T-D) interface to reflect the "reaction" of the distribution to the transmission system control. By using the response functions, only one or two iterations between the transmission system operator (TSO) and the distribution system operator(s) (DSO(s)) are required to attain a nearly optimal TDOPF solution. Numerical tests confirm that, relative to a typical decomposition method, the proposed method does not only enjoy a cheaper computational cost but is workable even when the objectives of the TSO and the DSO(s) are in distinct scales.

*Index Terms*--AC optimal power flow, distributed computation, distribution, transmission, response function.

## I. Introduction

Optimal power flow (OPF) is an important tool for optimizing power system operations. For interconnected power systems, the OPF tools installed in the control center of every system are often coordinated to achieve a globally optimal solution as well as to avoid the potential control conflicts between the control centers. This leads to a multi-area coordinated OPF problem which has been extensively studied in history. Abundant mathematical decomposition methods, e.g., APP [1], AND [2] and its derivations [3], ADMM [4] and HGD [5], have been attempted, network-equivalencing-based (NEB) methods [6] also tested.

Nevertheless, there are still imperfections in the above methods in terms of field applications. For example, these methods frequently entail tens or hundreds of the iterative updates of each area's OPF until converging, which inflicts huge communicational and computational burdens on every control center. The NEB methods may also deviate from the global optimum due to the errors in the equivalencing. Furthermore, the control centers to be coordinated may be concerned with diverse operational objectives, e.g., minimum generation cost or minimum electrical losses [7], and thus the multi-area coordinated OPF sometimes has to coordinate the distinct objectives that are in different scales. In this case, although a combination of a method of weight and one of the above decomposition methods is simple and usually employed, different choices of the weight may lead to dissimilar optimal solutions, which is also a confusing issue for system operators.

A special type of the multi-area coordinated OPF is the transmission-distribution-coupled AC OPF (TDOPF) [5] that a transmission system operator (TSO) is coordinated with one or several distribution system operators (DSOs) to achieve a globally optimal operating point, so the above methods and their defects both apply to the TDOPF. In fact, it may more frequently face the above different-scale-objective issue because the operational responsibility and the objective of a TSO usually differ from those of a DSO [8]. In order to overcome the above challenges in the TDOPF, it is necessary to present a new coordination method that both enjoys a cheaper computational cost and is capable of handling the different-scale-objective issue.

By reviewing the interaction features of an integrated transmission-distribution (T-D) power system as stated in [5], we notice that, currently a distribution system is almost always connected to a transmission system at one transmission bus (which is physically a distribution station and will be referred to a T-D interface in the remainder) and the voltage in the T-D interface is managed by the TSO and the power injections in the interface is supervised by the DSO. These facts imply a chance of employing the response functions of the power injections with respect to the voltage change in the T-D interface to fulfil a computationally cheap solution to the TDOPF. Hence, a response-function-based coordination method is proposed in this paper to solve the TDOPF, and we found that only one or two iterations between the TSO and the DSO(s) are required to attain a feasible solution which is also

This work was supported in part by the China Postdoctoral Science Foundation under Grants 2016M600091 and 2017T100078 and in part by the Foundation for Innovative Research Groups of the National Natural Science Foundation of China under Grant 51621065.

often close to the optimum. Furthermore, relative to the recently published method in [9], our method has two major advantages: 1) it involves T-D interface voltage feasible constraints to ensure that the calculation process will not break off due to an infeasible DSO's OPF problem; and 2) it addresses the different-scale-objective issue.

The remainder of this paper will be arranged as follows. In Section II, preliminaries such as the analysis of the iteration of the TSO and the DSO(s) are presented. In Section III, the proposed method is delineated with its features discussed. Case studies are shown in Section IV to verify the advantages of the proposed method. Finally, conclusions are drawn in Section V.

## II. PRELIMINARIES

### A. Analysis of Interaction of TSO and DSO

Fig. 1 shows a typical integrated T-D power system in which one transmission and two distribution systems are interconnected by two T-D interfaces. The interface is also referred to as the boundary of the TSO and the DSO. Since the voltage and power injection in the T-D interface are usually managed by the TSO and the DSO respectively [5], the corresponding optimal voltage (denoted by $V_B \angle \theta_B$ in Fig. 2) and the power injections (denoted by $P_B+jQ_B$ in Fig. 2) are determined by the TSO's OPF (T-OPF) and the DSO's OPF (D-OPF), respectively. Moreover, since a distribution system is usually connected to the transmission by one T-D interface, the voltage phase angle in the interface can always be taken as zero in the D-OPF formulation regardless of the specific value of $\theta_B$ produced by the T-OPF. Therefore, $V_B$, $P_B$ and $Q_B$ work as the coupling variables in the interaction of the TSO and the DSO(s), and this is illustrated in Fig. 2.

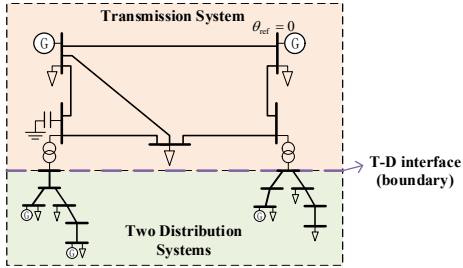

Figure 1. Diagram of a typical integrated T-D system

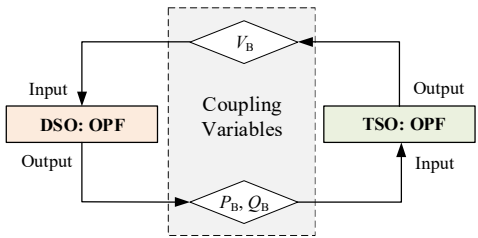

Figure 2. Illustration of the interaction of the TSO and the DSO

With a further insight into the interaction shown in Fig. 2, we also notice that, the optimum of the T-OPF is affected by the input parameters $P_B$ and $Q_B$ that are determined by the D-OPF, and meanwhile the optimum of the D-OPF is affected by the input parameter $V_B$ that is determined by the T-OPF. In this sense, if the response function of $P_B$ and $Q_B$ with respect to $V_B$ can be well evaluated and imparted to the TSO, then the (local) optimum of the T-OPF will also be at least locally optimal for the integrated T-D power system. This inspires us to propose the response-function-based coordination method that will be presented in Section III.

In addition, as pointed out in [5], the operational objectives of the TSO and the DSO are often distinct. For example, the former may intend to minimize the electrical losses while the latter would like to minimize the distributed generation (DG)'s cost in its own area. In this case, the TSO's and the DSO's objectives are in different scales and the different-scale-objective issue thus emerge. This issue will also be addressed by the proposed method.

### B. Formulations of TSO's and DSO's OPF

The compact formulations of the T-OPF and the D-OPF are presented in (1) and (2), respectively. AC OPF formulations are involved herein for general purpose:

$$\text{T-OPF:} \quad \begin{aligned} &\min_{\mathbf{x}_T,\mathbf{u}_T} c_T\left(\mathbf{x}_T,\mathbf{u}_T,\mathbf{P}_B,\mathbf{Q}_B\right) \\ &s.t. \begin{cases} \mathbf{f}_T\left(\mathbf{x}_T,\mathbf{u}_T,\mathbf{P}_B,\mathbf{Q}_B\right)=\mathbf{0} \\ \mathbf{g}_T\left(\mathbf{x}_T,\mathbf{u}_T,\mathbf{P}_B,\mathbf{Q}_B\right)\geq\mathbf{0} \\ \mathbf{x}_T \text{ includes } \mathbf{V}_B \end{cases} \end{aligned} \quad (1)$$

where the optimization variable vectors $\mathbf{x}_T$ and $\mathbf{u}_T$ represent the buses' voltages and control variables in a transmission system, respectively, and by definition the T-D interface voltage magnitude vector $\mathbf{V}_B$ is naturally a part of $\mathbf{x}_T$; the input parameter vectors $\mathbf{P}_B$ and $\mathbf{Q}_B$ comprise the active and reactive power injections in every T-D interface, respectively; the functions $c_T$, $\mathbf{f}_T$ and $\mathbf{g}_T$ are the operational objective, the AC power flow constraint and the inequality constraints (e.g., constraints of the line flow, bus voltage and generator output) for a TSO, respectively.

$$\text{D-OPF:} \quad \begin{aligned} &\min_{\mathbf{x}_D,\mathbf{u}_D} c_D\left(\mathbf{x}_D,\mathbf{u}_D,V_B\right) \\ &s.t. \begin{cases} \mathbf{f}_D\left(\mathbf{x}_D,\mathbf{u}_D,V_B\right)=\mathbf{0}, \ \mathbf{g}_D\left(\mathbf{x}_D,\mathbf{u}_D,V_B\right)\geq\mathbf{0}, \\ P_B=h_{PB}(\mathbf{x}_D,V_B), \quad Q_B=h_{QB}(\mathbf{x}_D,V_B) \end{cases} \end{aligned} \quad (2)$$

where the optimization variable vectors $\mathbf{x}_D$ and $\mathbf{u}_D$ represent the nodal voltages and control variables in a distribution system, respectively; the functions $c_D$, $\mathbf{f}_D$ and $\mathbf{g}_D$ denote the objective, the AC power flow constraint and the inequality constraints for a DSO, respectively; the input parameter $V_B$ is the T-D interface voltage magnitude and the output $P_B$ and $Q_B$ are determined by the power equations in the T-D interface, which are denoted by $h_{PB}$ and $h_{QB}$, respectively. Notably, the respective dimension of $V_B$, $P_B$ and $Q_B$ in (2) is one since one T-D interface is for one distribution system as analyzed above.

More detailed formulations of the T-OPF and D-OPF can be found in [5]. To save space, only the above compact formulations are presented here and will used in the remainder.

## III. RESPONSE-FUNCTION-BASED COORDINATION METHOD

### A. General Idea

The above preliminaries indicate that, if the response functions of $P_B$ and $Q_B$ with respect to $V_B$ can be accurately evaluated, then the (local) optimum of the TDOPF can be obtained by solving the T-OPF and the D-OPF only once, so the repeated iterative updates of the OPF are avoided. In fact, because of the non-convexity of AC OPF formulations, it is usually more practicable to make an approximate, rather than accurate, estimation of the response function, but we can still expect that a nearly optimal solution to the TDOPF will be attained with very few iterative updates of the T-OPF and the D-OPF. Hence, the primary step in this method is to make a fair estimate of the response function of $P_B$ and $Q_B$ with respect to $V_B$ in a T-D interface. Additionally, we should also determine the domain of the response function so that any $V_B$ in the domain ensures the feasibility of the corresponding D-OPF, which is a precondition for the proposed method producing a reasonable solution.

Therefore, the general idea of this method is **1)** first to create T-D interface voltage feasible constraints; and **2)** then to evaluate the response functions and add them into the T-OPF model to obtain the (local) optimum $(\mathbf{x}_T, \mathbf{u}_T)$ which are also of acceptable optimality in terms of an integrated T-D system; and **3)** to obtain $(\mathbf{x}_D, \mathbf{u}_D)$ by solving (2) that is with $V_B$ assigned with its counterpart in $\mathbf{x}_T$. The details will be explained below.

### B. T-D Interface Voltage Feasible Constraint

Conventionally, the upper and lower limits regarding $V_B$ are presumed to be offline predetermined, e.g., 0.95 p.u. and 1.05 p.u., respectively. Sometimes, however, even if these predetermined limits are involved in the T-OPF, the produced $V_B$ is still possible to induce an infeasible D-OPF, especially when there is considerable fluctuating DG's integration. For example, in the case of abundant DG output, the upper limit of $V_B$ sometimes has to be decreased from the offline set value to a smaller one so that any $V_B$ satisfying the new limit ensures that the solution to the corresponding D-OPF will not violate the upper nodal voltage limit or other constraints. Therefore, the T-D interface voltage feasible constraint, which is a constraint regarding $V_B$ to ensure the D-OPF is feasible, should be first created to bound the domain of the response function.

The voltage feasible constraint in one T-D interface is formulated as $\underline{V}_B^f \leq V_B \leq \overline{V}_B^f$, where the lower and upper limits $\underline{V}_B^f$ and $\overline{V}_B^f$ should be computed via the optimization problems in (3) and (4), respectively, which are both solved by the DSO:

$$\overline{V}_B^f = \max_{\mathbf{x}_D, \mathbf{u}_D, V_B} V_B \qquad \underline{V}_B^f = \min_{\mathbf{x}_D, \mathbf{u}_D, V_B} V_B$$
$$s.t. \begin{cases} \mathbf{f}_D(\mathbf{x}_D, \mathbf{u}_D, V_B) = \mathbf{0} \\ \mathbf{g}_D(\mathbf{x}_D, \mathbf{u}_D, V_B) \geq \mathbf{0} \\ \underline{V}_B \leq V_B \leq \overline{V}_B \end{cases}, (3) \quad s.t. \begin{cases} \mathbf{f}_D(\mathbf{x}_D, \mathbf{u}_D, V_B) = \mathbf{0} \\ \mathbf{g}_D(\mathbf{x}_D, \mathbf{u}_D, V_B) \geq \mathbf{0} \\ \underline{V}_B \leq V_B \leq \overline{V}_B \end{cases}, (4)$$

where $\underline{V}_B$ and $\overline{V}_B$ are the conventional upper and lower limits of $V_B$, e.g., 0.95 p.u. and 1.05 p.u., respectively.

Obviously, we have $\underline{V}_B \leq \underline{V}_B^f \leq \overline{V}_B^f \leq \overline{V}_B$. In fact, the rightmost inequality is strict in the numerical test in Section IV, indicating that this voltage feasible constraint successfully weeds out the range of VB that may induce infeasible D-OPF.

### C. Response Function in a T-D Interface

To make a fair and easy-implementing estimate of the response function, a perturbation and fitting method is employed here.

First, a DSO solves the problem in (5) where $P_B$, $Q_B$ and $V_B$ are all regarded as optimization variables. Let $V_B^*$, $P_B^*$ and $Q_B^*$ denote the optimal solution.

$$\min_{\mathbf{x}_D, \mathbf{u}_D, V_B} c_D(\mathbf{x}_D, \mathbf{u}_D, V_B)$$
$$s.t. \begin{cases} \mathbf{f}_D(\mathbf{x}_D, \mathbf{u}_D, V_B) = \mathbf{0}, \ \mathbf{g}_D(\mathbf{x}_D, \mathbf{u}_D, V_B) \geq \mathbf{0} \\ P_B = h_{PB}(\mathbf{x}_D, V_B), \quad Q_B = h_{QB}(\mathbf{x}_D, V_B) \\ \underline{V}_B^f \leq V_B \leq \overline{V}_B^f \end{cases}, (5)$$

Second, the DSO computes $V_1 = \max\{(1-\alpha)V_B^*, \underline{V}_B^f\}$ and $V_2 = \min\{(1+\alpha)V_B^*, \overline{V}_B^f\}$ with a parameter $\alpha$ whose value is predetermined, e.g., 0.01. The DSO then solves two corresponding problems that are formulated in (6) and (7). Let $P_{B,1}, Q_{B,1}$ and $P_{B,2}, Q_{B,2}$ denote the optimal values of the power injections in the T-D interface in (6) and (7), respectively.

$$\min_{\mathbf{x}_D, \mathbf{u}_D} c_D(\mathbf{x}_D, \mathbf{u}_D, V_1) \qquad \min_{\mathbf{x}_D, \mathbf{u}_D} c_D(\mathbf{x}_D, \mathbf{u}_D, V_2)$$
$$s.t. \begin{cases} \mathbf{f}_D(\mathbf{x}_D, \mathbf{u}_D, V_1) = \mathbf{0} \\ \mathbf{g}_D(\mathbf{x}_D, \mathbf{u}_D, V_1) \geq \mathbf{0} \end{cases}, (6) \quad s.t. \begin{cases} \mathbf{f}_D(\mathbf{x}_D, \mathbf{u}_D, V_2) = \mathbf{0} \\ \mathbf{g}_D(\mathbf{x}_D, \mathbf{u}_D, V_2) \geq \mathbf{0} \end{cases}. (7)$$

Next, the DSO applies a piecewise-linear fitting method, e.g., the one provided by MATLAB, to the data $\{(V_B^*, P_B^*), (V_1, P_{B,1}), (V_2, P_{B,2})\}$ and $\{(V_B^*, Q_B^*), (V_1, Q_{B,1}), (V_2, Q_{B,2})\}$ to obtain a piecewise-linear function of $P_B$ with respect to $V_B$, which will be used as an approximate response function in the domain $V_1 \leq V_B \leq V_2$. The expression is shown below:

$$P_B = \begin{cases} a_{P,1} V_B + b_{P,1}, & V_1 \leq V_B \leq V_B^* \\ a_{P,2} V_B + b_{P,2}, & V_B^* \leq V_B \leq V_2 \end{cases}, \quad (8)$$

where $a_{P,1}$, $b_{P,1}$, $a_{P,2}$ and $b_{P,2}$ are the coefficients that are the output of the fitting method.

Similarly, an approximate response function of $Q_B$ with respect to $V_B$ in the domain $V_1 \leq V_B \leq V_2$ is obtained and shown below:

$$Q_B = \begin{cases} a_{Q,1}V_B + b_{Q,1}, & V_1 \leq V_B \leq V_B^* \\ a_{Q,2}V_B + b_{Q,2}, & V_B^* \leq V_B \leq V_2 \end{cases}, \quad (9)$$

where $a_{Q,1}$, $b_{Q,1}$, $a_{Q,2}$ and $b_{Q,2}$ are the coefficients that are the output of the fitting method.

### D. Complete Procedures of the Method

The complete procedures contain the following steps:

**Step 1**: A DSO creates the constraint $\underline{V}_B^f \leq V_B \leq \overline{V}_B^f$ by solving the problems in (3) and (4).

**Step 2**: With a predetermined value of $\alpha$, the DSO solves the problems in (5), (6) and (7), and then obtains the approximate response functions of $P_B$ and $Q_B$ with respect to $V_B$, as are described in (8) and (9), respectively.

**Step 3**: The DSO imparts to the TSO the constraint $V_1 \leq V_B \leq V_2$ and the response functions in (8) and (9). The TSO then solves a modified T-OPF as shown in (10). This problem can be solved by current available solvers, e.g., the BONMIN provided by the YALMIP [10]. Let $V_{B,T}$ denote the optimal value of $V_B$ in (10) and the TSO then imparts $V_{B,T}$ to the DSO.

$$\min_{\mathbf{x}_T, \mathbf{u}_T, P_B, Q_B} c_T(\mathbf{x}_T, \mathbf{u}_T, P_B, Q_B)$$

$$s.t. \begin{cases} \mathbf{f}_T(\mathbf{x}_T, \mathbf{u}_T, P_B, Q_B) = \mathbf{0} \\ \mathbf{g}_T(\mathbf{x}_T, \mathbf{u}_T, P_B, Q_B) \geq \mathbf{0} \\ \mathbf{x}_T \text{ includes } V_B, \quad V_1 \leq V_B \leq V_2 \\ P_B, Q_B, V_B \text{ subject to (8), (9)} \end{cases} \quad (10)$$

**Step 4**: The DSO solves (2) with $V_B = V_{B,T}$. Let $P_{B,D}$ and $Q_{B,D}$ denote the optimal values of $P_B$ and $Q_B$ in this problem.

**Step 5**: Compare $P_{B,D}$ and $Q_{B,D}$ and their counterparts that are evaluated using (8) and (9) with $V_B = V_{B,T}$.

It the difference is sufficiently small (e.g., $10^{-4}$ in the tests in Section IV), terminate the algorithm and the solution obtained in the above steps can be regarded as a nearly optimal solution to the TDOPF. That is because the small difference indicates that the power mismatches in the T-D interface, which are resulted from the solutions in (10) and (2), are negligible.

Otherwise, if there is a notable difference, then invoke Step 6 to update the solution.

**Step 6**: This step is to eliminate the power mismatches in the T-D interface. To this end, the TSO should solve (11) to update the solution. Obviously, per the constraints in (11), there will be no power mismatches in the T-D interface. After that, terminate the algorithm.

$$\min_{\mathbf{x}_T, \mathbf{u}_T} c_T(\mathbf{x}_T, \mathbf{u}_T, P_B, Q_B)$$

$$s.t. \begin{cases} \mathbf{f}_T(\mathbf{x}_T, \mathbf{u}_T, P_B, Q_B) = \mathbf{0} \\ \mathbf{g}_T(\mathbf{x}_D, \mathbf{u}_D, P_B, Q_B) \geq \mathbf{0} \\ P_B = P_{B,D}, Q_B = Q_{B,D}, V_B = V_{B,T} \end{cases} \quad (11)$$

### E. Further Discussions

**Remark 1 (multiple-DSO case)**: The above procedures also work for the case where multiple distribution systems are connected to one transmission system. In this case, Steps 1–4 and 6 should be performed by every DSO in a distributed fashion.

**Remark 2 (different-scale-objective issue)**: The different-scale-objective issue is addressed by the proposed method in the sense that summing up the distinct objectives of the TSO and the DSO(s), which is almost invariably required in a mathematical decomposition method, is subtly avoided as shown above.

**Remark 3 (feasible D-OPF)**: Instead of the original $\underline{V}_B \leq V_B \leq \overline{V}_B$, a tighter constraint $V_1 \leq V_B \leq V_2$ is used to ensure the feasibility of the DSO-P. Since $\underline{V}_B^f \leq V_1 \leq V_2 \leq \overline{V}_B^f$ and $\underline{V}_B^f$ and $\overline{V}_B^f$ are the maximum and minimum values of $V_B$ with which a solution can be found for the DSO-P (please review (2), (3) and (4)), $V_{B,T}$ solved using (10) will ensure the feasibility of the DSO-P. Therefore, the D-OPF is feasible in the calculation process.

**Remark 4 (optimality)**: Per parametric programming theory [11], the first-order response function of $P_B$ (or $Q_B$) with respect to $V_B$ should be a piecewise-affine function. However, it is often very expensive to evaluate the function accurately. Hence, a perturbation and fitting method is employed here to make an approximate estimation. Despite the inevitable errors in the approximation process in (5), (6) and (7), the numerical test in Section IV shows that the proposed method attains a nearly optimal solution.

Moreover, the parameter $\alpha$ in Step 3 is used to compromise the optimality of the T-OPF and the D-OPF. For example, $\alpha = 0.01$ means that $V_{B,T}$ is allowed to deviate no more than $\pm 1\%$ from $V_B^*$, implying that the DSO would sacrifice some of its optimality to allow the TSO to have a larger feasible region in (10) and, possibly, a better solution. The value of $\alpha$ can be determined by prior negotiations between the TSO and DSO(s).

**Remark 5 (very few iterations)**: This method requires no more than two iterative updates of the T-OPF and six for the D-OPF. If an "iteration" is defined as once data exchange between the TSO and DSO conventionally, then this method requires no more two iterations between the TSO and the DSO(s).

## IV. CASE STUDIES

We tested the proposed method on two integrated T–D systems. The first, called T14D1, is made by connecting an IEEE 14-bus system to a 38-node distribution system, and the second is a T14D4 system with one transmission system and four distribution systems [5]. In the simulation, the specific formulations of the T-OPF and the D-OPF are the same as [5], which includes generators' output as control variables and the AC power flow equations and the line flow/bus voltage/generator output inequality constraints, with the following exceptions: 1) $c_T$ aims to minimize the power loss of the transmission network with the active power of all transmission generators (TGs) other than the slack generator assumed to be 30 MW; and 2) $c_D$ aims to minimize the sum of the DG's generation cost and the cost of buying power from the transmission side, and the cost is set as 40 \$/MWh as per the average marginal price of the IEEE 14-bus system.

First, the proposed method is compared to a centralized TDOPF model which involves all the constraints in the D-OPF and the T-OPF and it has a weighted objective function $\omega_T c_T + \omega_D c_D$ where $\omega_T$ and $\omega_D$ are the weights. Table I presents the relative results. The 2nd and 3rd rows indicate that the optimal solution of a centralized model notably varies with the different weights used in the objective function. In fact, because the objectives of the TSO and the DSOs are in different scales, it is usually difficult to determine which of the two optimums is the "right" one. In contrast, the proposed method addresses this confusing issue and attains a compromised coordination solution (please see the columns 6 and 7 in Table I).

Second, as for the optimality, Table II lists the weighted objectives evaluated with the solution obtained by the proposed method as well as the optimal solutions to the two centralized models with objectives ($c_T+c_D$) and ($20c_T+c_D$). The solutions to the central.($c_T+c_D$) or ($20c_T+c_D$) model naturally performs best in minimizing the function ($c_T+c_D$) or ($20c_T+c_D$), but the proposed method also successfully achieves a nearly optimal solution with very few iterations (please see Table III) and thus saves considerable computational time.

Third, Table III compares the required iterations of the proposed method and the APP method [1] that is used to distributedly solve a centralized TDOPF model. Notably, the APP is tested on both central.($c_T+c_D$) and central.($20c_T+c_D$) models to reflect the impact of weights on the iterations. The result shows that the proposed method needs only one or two iterations, whereas the APP always requires dozens of iterations between the TSO and the DSOs regardless of the weights. Hence, in the case of a restricted communication environment, the proposed method is remarkably preferable to the decomposition method. Furthermore, as there is lack of the T-D voltage feasible constraint, the method in [9] failed to produce a solution on the T14D4 system as one D-OPF problem became infeasible in the calculation, which is due to an inappropriate $V_B$ sent by the TSO that precisely falls outside the domain which ensures feasible D-OPF.

TABLE I. COMPARISON WITH THE OPTIMAL SOLUTIONS TO WEIGHTED CENTRALIZED TDOPF MODELS ON T14D1

| Items | $P_{TG}$ (MW) | $P_{DG}$ (MW) | $P_B$ (MW) | $Q_B$ (Mvar) | $c_T$ (MW) | $c_D$ (\$) |
|---|---|---|---|---|---|---|
| Central.[a] ($c_T+c_D$) | [142.45; 30; 30; 30; 30] | [1; 1; 2; 1.5; 0] | 6.764 | 2.119 | 5.686 (3rd best) | 472.66 (best) |
| Central. ($20c_T+c_D$) | [140.29; 30; 30; 30; 30] | [1; 1; 2; 1.5; 2] | 4.757 | 3.211 | 5.533 (best) | 474.41 (3rd best) |
| Proposed method ($\alpha$ = 0.01) | [142.43; 30; 30; 30; 30] | [1; 1; 2; 1.5; 0] | 6.770 | 3.321 | 5.666 (2nd best) | 473.70 (2nd best) |

a. Central. ($c_T+c_D$) or ($20c_T+c_D$) represents a centralized TDOPF model that includes the constraints in (1) and (2) and has the objective $c_T+c_D$ or $20c_T+c_D$.

TABLE II. OPTIMALITY VERIFICATION

| Weighted Objectives Evaluated with Different Solutions | T14D1 | | T14D4 | |
|---|---|---|---|---|
| | $c_T+c_D$ | $20c_T+c_D$ | $c_T+c_D$ | $20c_T+c_D$ |
| Optimum of Central. ($c_T+c_D$) | 478.346 | 586.38 | 124.630 | 189.120 |
| Optimum of Central. ($20c_T+c_D$) | 479.943 | 585.07 | 124.720 | 188.877 |
| Solution of the proposed method ($\alpha$ = 0.01) | 479.366 | 587.02 | 124.634 | 189.122 |

TABLE III. ITERATIONS BETWEEN THE TSO AND DSO(S)

| Methods | T14D1 | | T14D4 | |
|---|---|---|---|---|
| | Central. ($c_T+c_D$) | Central. ($20c_T+c_D$) | Central. ($c_T+c_D$) | Central. ($20c_T+c_D$) |
| APP: distributedly solves a centralized TDOPF model | 17 | 20 | 85 | 49 |
| Proposed method ($\alpha$ = 0.01) | 1 | | 2 | |

## V. CONCLUSIONS

This paper proposes employing the response functions of the power injections with respect to the voltage magnitude in the T-D interface to solve the TDOPF. Since the response function reflects the "reaction" of the distribution to the transmission system control, this method will attain a nearly optimal solution with only one or two iterations between the TSO and the DSO(s). Moreover, the method is also workable even when the objectives of the TSO and the DSO(s) are in distinct scales. Case studies confirm that the proposed method is computationally cheaper relative to a typical decomposition method and thus preferable when there is a restricted communication environment or a different-scale-objective issue in the TDOPF.